\documentclass[singlecolumn,preprint,12pt,3p]{elsarticle}

\usepackage{amssymb}

\journal{}

\begin{document}

\begin{frontmatter}

\title{Study of Bose-Einstein condensation using generalized canonical partition function}

\author{Sarath.R}
\address{Department of Applied Physics, Indian School of Mines, Dhanbad, Jharkhand, India-826004}
\ead{sarath91@live.com}
\author{P. C. Vinodkumar}
\address{Department of Physics, Sardar Patel University,Vallabh Vidyanagar, India-388120}
\ead{pothodivinod@yahoo.com}
\begin{abstract}
We open a new discussion of generalized canonical partition function in standard statistical mechanics and apply it for the study of Bose-Einstein condensation. We discuss the possible cases for the generalized canonical partition function and arrives at a conclusion that the system of trapped bose gas will not be existing at absolute zero. We analyse the present study with an experimental result and point out the general difficulties in the analyses of experimental observations, which can possibly suppress the effect of generalized canonical partition function over standard canonical partition function. We mention that the experimental studies with ideal condensates at absolute zero with an unbiased approach towards the traditional Bose-Einstein condensation theory can bring out the effect of generalized canonical partition function.
\end{abstract}


\end{frontmatter}


\section{Introduction}
Bose-Einstein condensation (BEC) is regarded as one of the most important theoretical predictions by Einstein \cite{EINSTEIN} from Bose's paper on Plank's formula \cite{BOSE}. It was London who recognized it as an important physical concept by connecting it with the problem of liquid helium \cite{LON}. However, the first experimental realization of Bose-Einstein condensation came only after a very long time at ultra-cold temperature scale in alkali gases \cite{ANDSN,BRA,DAVIS}. Initially the experimental BEC studies were concentrated on the alkali gases, later BEC was observed in quasi particles like, excitons and magnons \cite{EXC1,EXC2,EXC3,EXC4,EXC5} also. Quite recently, the BEC of light or specifically photon, the most fundamental boson based on which all the quantum theories are formulated has been reported \cite{PHT,PHT1}. The BEC of light promises an exploration in the study of ideal bose gases \cite{PHT,PHT2}.\\

Inter-particle interactions are important in the study of BEC. The non-interacting bose gases may appear trivial from a theoretical point of view, but from an experimental point of view it is very difficult to create an ideal BEC due to the non vanishing scattering length of the bosons. However, the ideal condensate and the interaction tunable condensates are gaining a lot of attractions in recent experimental studies \cite{TUNK,SAR}. An ideal Bose-Einstein condensate can be used to search new physics in the phenomena of Anderson Localization and matter-wave interferometry \cite{ANDLOC,ANDLOC1,MAT1,MAT2,MAT3} and the interaction tunable Bose-Einstein condensate can be helpful for the study of Heisenberg-limited interferometry \cite{HSB1,HSB2,HSB3}. However, the increasing experimental studies on non-interacting bose gas does not diminish the importance of interacting bose gas. Recent study given in Ref.\cite{BLOCH} prepared a negative temperature scale for the motional degrees of freedom, using an attractively interacting ensemble of ultra-cold bose gas.\\

Another important factor in the study of the BEC is the trapping potential. Usual graduate level textbooks discuss the condensation of free particles that is in a 3D box potential, more specifically spatially uniform bose gas. However, for the experimental realization, harmonic potential is the most favorable trapping potential and thus the confining bose gas would be non uniform with no translational symmetry \cite{UNI}. The basic problem is that the non-uniformity  of the confined  bose gas may suppress the underlying physics. To resolve this problem up to an extent, the studies concentrated on the local density approximation and the small central portion of the trapped bose gas \cite{UNI,UNI1,UNI2,UNI3,UNI4,UNI5,UNI6,UNI7,UNI8,UNI9}. The recent creation of the Bose-Einstein condensation in a three dimensional quasi uniform optical box trap is a step forward towards the study of fundamental studies of BEC, since it can replace the studies concentrated on the local density approximation and the small central portion of the trapped bose gas \cite{UNI}. These studies show an unusual way of reverse order in the experimental and theoretical studies of BEC. Those studies which are easy in a theoretical perspective are difficult in experimental studies and viceversa.\\

Theoretically,  BEC is usually discussed in the grand canonical ensemble \cite{PATRIA,CN1,KERTLE}. However, there is a major problem of huge fluctuations in the ground state occupation number with the grand canonical ensemble for a non-interacting case \cite{CN1}. Ref. \cite{CN2}, indicates that, all the three ensembles, ie. microcanonical, canonical and grand canonical are valid to use for the study of BEC. However, for a non-interacting case, the ground state fluctuations for three different ensembles can be represented in the order, $microcanonical<canonical<<grand-canonical$ \cite{CN3,CN4,CN5}. The presence of inter-particle interactions reduces the fluctuations in the ground state occupation in grand canonical ensemble \cite{CN6}, but for a non-interacting case the grand canonical ensemble appears not as an appropriate choice.\\

On the other hand Supersymmetry is a theoretical formulation which was supposed to answer some fundamental questions in the high energy physics. Even though Supersymmetry (SUSY) has not been observed in nature, it's mathematical formulations find a lot of applications in other fields of Physics including quantum mechanics. For instance, Supersymmetric quantum mechanics (SUSY QM) \cite{WITT,FREED,COOPER} provides a better understanding about the analytically solvable potential problems \cite{COOPER}. A recent analogues study of SUSY QM on statistical mechanics \cite{SUSYSM} shows the canonical partition function of ordinary statistical mechanics as a special case of a generalized canonical partition function. This resembles the hidden variable theory of wave function in quantum mechanics \cite{GRIFFITH}. The generalized canonical partition function indicate that the canonical partition function of the ordinary statistical mechanics does not provide the whole picture, in fact there is a hidden term which may not be detected using the normal statistical experiments. However, the generalized partition function is not just valid only in the SUSY context, it is also equally valid in the frame of standard statistical mechanics. It is because of the mathematical similarity of SUSY QM with the construction of generalized canonical partition function, we could recognize the hidden term which is not accounted in the standard statistical mechanics. Thus it is of prime importance to see the possible effects of generalized canonical partition function over the standard canonical partition function. In the present study, we aim to see the implications of the generalized canonical partition function in BEC, since it is an important low temperature quantum statistical phenomenon. However, for the present study we consider a non-interacting bose gas, to make the system simpler, confined in a 3D harmonic trap, because of it's experimental abundance and importance.\\

The present study is organized in five parts. In the second part we shall briefly review the construction of generalized canonical partition function in the frame of standard statistical mechanics. In the third section we will discuss the BEC in 3D harmonic potential. In the fourth section we will analyse our present study with an experimental data. In the fifth and last part we draw important conclusions of the present study.
\section{Generalized canonical partition function}
We begin our discussion with a brief review of construction of the generalized canonical partition function \cite{SUSYSM} in the frame of standard statistical mechanics.\\

By standard statistical mechanics, the internal energy can be expressed as,
\begin{equation}\label{eq:1.1}
U(T)= k_{B} T^{2} \frac{d}{dT}ln Z(T)
\end{equation}
Where $k_{B}$ is the Boltzmann constant, $T$ is the temperature and $Z(T)$ is the standard canonical partition function as a function of temperature.\\

With the transformation $x=-1/k_{B}T$, the internal energy can be written as,
\begin{equation}\label{eq:1.2}
U(x)= \frac{Z'(x)}{Z(x)}
\end{equation}
Where the prime represent the derivative with respect to $x$.\\

Thus we can form Ricatti equation as,
\begin{equation}\label{eq:1.3}
U^{2}(x)+U'(x)= \frac{Z''(x)}{Z(x)}
\end{equation}
The solution of the Ricatti equation (\ref{eq:1.3}) will give a generalized internal energy, $U_{g}(x)$ as \cite{SUSYSM},
\begin{equation}\label{eq:1.4}
U_{g}(x)=\frac{d}{dx}\ln(Z(x))+\frac{d}{dx}ln(\gamma+b\int\frac{dx}{Z^{2}(x)})
\end{equation}
Where $\gamma$ and $b$ are constants arise from the solution of Riccati equation \cite{SUSYSM}.\\

Thus we can extract a generalized canonical partition function, $Z_{g}(x)$  from the $U_{g}(x)$, as,
\begin{equation}\label{eq:1.5}
Z_{g}(x)=Z(x)(\gamma+b\int\frac{dx}{Z^{2}(x)})
\end{equation}
We rewrite the generalized canonical partition function as,
\begin{equation}\label{eq:1.6}
Z_{g}(x)=Z(x)(1+\beta\int\frac{dx}{Z^{2}(x)})
\end{equation}
Where $\beta=b/\gamma$.
Thus with the standard statistical relations, we can construct the generalized canonical partition function. Now from equation (\ref{eq:1.6}), we can see that the generalized canonical partition function has an extra term along with the standard canonical partition function. The Supersymmetric (SUSY) connection comes because the Ricatti equation (\ref{eq:1.3}) is an analogous equation for the SUSY quantum mechanical relation, with internal energy analogous to Witten's super potential and canonical partition function analogous to the ground state wave function \cite{SUSYSM,COOPER}. However, with this analogous relation, one cannot simply underestimate the importance of generalized canonical partition function in the standard statistical mechanics.\\

Let us take the case of a three dimensional harmonic potential. By standard statistical mechanics the canonical partition function can be expressed, semiclassically as,
\begin{equation}\label{eq:1.7}
Z(x)=\int_{0}^{\infty}g(E) \exp(xE) dE
\end{equation}
Where $g(E)$ is the density of states. The density of states for a 3D harmonic potential, is well known and can be written as,
\begin{equation}\label{eq:1.8}
g(E)=\frac{E^{2}}{2 \hbar^{3}\omega_{x}\omega_{y}\omega_{z}}
\end{equation}
Where $\omega_{x}$, $\omega_{y}$ and $\omega_{z}$ are the frequencies along $x$, $y$ and $z$ directions respectively. Thus the standard canonical partition function can be obtained as,
\begin{equation}\label{eq:1.9}
Z(x)=-\frac{1}{(x\hbar)^{3}\omega_{x}\omega_{y}\omega_{z}}
\end{equation}
Hence the generalized partition function can be obtained as,
\begin{equation}\label{eq:1.10}
Z_{g}(x)=Z(x)(1+\frac{\beta x^{7}\hbar^{6}\omega_{x}^{2}\omega_{y}^{2}\omega_{z}^{2}}{7})
\end{equation}
As a function of temperature, the generalized canonical partition function for a 3D harmonic potential can be expressed as,
\begin{equation}\label{eq:1.11}
Z_{g}(T)=\frac{(k_{B}T)^{3}}{\hbar^{3}\omega_{x}\omega_{y}\omega_{z}}-\beta\frac{\hbar^{3}\omega_{x}\omega_{y}\omega_{z}}{7k_{B}^{4}T^{4}}
\end{equation}
Where the standard canonical partition function given in equation (\ref{eq:1.9}) is written explicitly as a function of temperature.\\

From equation (\ref{eq:1.11}) it is clear that the constant $\beta$ has the dimension of energy. However, $\beta$ cannot be a temperature dependent term since it is a ratio of constants arising from the solution of the Ricatti equation (\ref{eq:1.3}).\\

One interesting thing to notice is that the extra term in the generalized partition function has temperature dependence in the denominator. This extra term will dominate the standard canonical partition function in the very low temperature regime, especially for $T\rightarrow0$. Thus the ultra-cold phenomena like BEC would be a perfect choice to study it's effect.
\section{Bose-Einstein condensation in 3D harmonic potential}
Here we study the Bose-Einstein condensation of finite number of particles in a 3D harmonic potential by using the generalized canonical partition function which we have obtained in the previous section using the semiclassical method. Here we concentrate on the contribution from the excited states, hence the semiclassical method will give good approximate results \cite{PETHIK}.
With canonical partition function we can express the number of particles ($N$) in the 3D harmonic potential as \cite{HOL,KERTLE},
\begin{equation}\label{eq:1.12}
N=N_{0}+\textrm{g}_{3}(\textit{z})Z(T)
\end{equation}
Where $\textit{z}$ is the fugacity, $\textrm{g}_{n}$ is the Bose function and $N_{0}$ is the number of condensed particles. For temperatures $T\leq T_{c}$, the number of particles in the excited states, $N_{exc}$ is \cite{HOL,KERTLE,PATRIA},
\begin{equation}\label{eq:1.13}
N_{exc}=\zeta(3) Z(T)
\end{equation}
Where $\zeta(n)$ is the Riemann zeta function. However, we have seen that the standard canonical partition function, $Z(T)$  is only a special case of the generalized canonical partition function  $Z_{g}(T)$. Thus we replace $Z(T)$ in equation (\ref{eq:1.13}) by $Z_{g}(T)$ and obtain,
\begin{equation}\label{eq:1.14}
N_{exc}=\zeta(3)(\frac{(k_{B}T)^{3}}{\hbar^{3}\omega_{x}\omega_{y}\omega_{z}}-\beta\frac{\hbar^{3}\omega_{x}\omega_{y}\omega_{z}}{7k_{B}^{4}T^{4}})
\end{equation}\\

We can analytically deduce the expressions of generalized internal energy, $U_{g}(T)$ for a particle confined in a $3D$ harmonic potential as,
\begin{equation}\label{eq:1.15}
U_{g}=-4k_{B}T-\frac{49k_{B}^{8}T^{8}}{-7k_{B}^{7}T^{7}+\beta\hbar^{6}\omega_{x}^{2}\omega_{y}^{2}\omega_{z}^{2}}
\end{equation}\\

From equations (\ref{eq:1.14}) and (\ref{eq:1.15}) we can realize that the $\beta$ turns out to be an important factor. When $\beta=0$ all the relations will reduce to the case of usual BEC theory. However, $\beta\neq0$, deserves an analyses and this situation can have two different cases either $\beta>0$ or $\beta<0$.\\

\emph{\textsc{1}.When $\beta>0$}\\

For $\beta>0$, equation (\ref{eq:1.14}) suggest that , the number of bosons in the excited energy levels will be zero and thus form a fully condensed state at some positive temperature, $T_{0}= \frac{1}{k_{B}}(\frac{\beta \hbar^{6} \omega_{x}^{2}\omega_{y}^{2}\omega_{z}^{2}}{7})^{1/7}$ above absolute zero. However, from equation (\ref{eq:1.15}), one can easily see that the generalized internal energy $U_{g}(T)$ will be undefined at $T_{0}$ and for temperatures below $T_{0}$ the generalized internal energy will be negative. At absolute zero the generalized internal energy will become zero. Thus it shows that for $T\leq T_{0}$ the system becomes unphysical. This implies that the system with bosons trapped in the harmonic potential will no longer exist for $T\leq T_{0}$. This could be possibly because the trapping potential may break and the system may collapse immediately after forming a fully condensed state at $T_{0}$.\\

\emph{\textsc{2}.When $\beta<0$}\\

For $\beta<0$, equation (\ref{eq:1.14}) suggest that the number of bosons in the excited energy level will never be zero at any temperature and there will not be a fully condensed state even at absolute zero. However, the generalized internal energy, $U_{g}$ given in equation (\ref{eq:1.15}) will become zero at absolute zero and at temperature $T_{1}= 0.7890782986486072\frac{1}{k_{B}}( |\beta| \hbar^{6} \omega_{x}^{2}\omega_{y}^{2}\omega_{z}^{2})^{1/7}$. The first derivative of $U_{g}$ is zero at temperature $T= 0.5306184162532914\frac{1}{k_{B}}(|\beta| \hbar^{6} \omega_{x}^{2}\omega_{y}^{2}\omega_{z}^{2})^{1/7}$ and the second derivative is positive at this temperature. This means there is a minima in between absolute zero and $T_{1}$ and thus the generalized internal energy is negative in between absolute zero and $T_{1}$. This is an unphysical situation to internal energy be negative. Which implies the system of trapped boson is no longer exist below $T_{1}$ and as we have discussed in the $\beta>0$ case that this could be possibly because the trap may break at $T_{1}$, so that the system with the partially condensed bosons may collapse.
\section{Analyses with the experimental result}
We consider the experimental study of Bose-Einstein condensation by Mewes $et$ $al$ \cite{EXPARA} as a general representative of the experimental studies reported so far. At phase transition, the total number of sodium atoms confined in the dc magnetic trap with frequencies $\omega_{x}=2\pi\times18$, $\omega_{y}=2\pi\times320$ and $\omega_{z}=2\pi\times320$ is $N=15\times10^{6}$. The experimentally obtained transition temperature is $T_{c}=2\mu K$ and the critical peak density at $2\mu K$ is $n_{c}=1.5\times10^{14}$ $cm^{-3}$.\\

The transition temperature for the non-interacting bose gas confined in the 3D harmonic potential is given as \cite{HOL,KERTLE},
\begin{equation}\label{eq:1.16}
T^{0}_{c}=(\frac{N}{\zeta(3)})^{1/3}\frac{\hbar\overline{\omega}}{k_{B}}
\end{equation}
Where $\overline{\omega}=(\omega_{x}\omega_{y}\omega_{z})^{1/3}$. Equation (\ref{eq:1.16}) is accurate only at the thermodynamic limit and for the system with finite number of particles, the finite size effect will introduce corrections to the transition temperature \cite{HOL,KERTLE}. However, only the first order corrections for the finite size effect had been available in literature \cite{HOL,KERTLE} until the recent analytical derivation of higher order corrections \cite{SEC,JOSE}. The higher order corrections for the finite size effect leads to the argument that the Bose-Einstein condensation is not a well defined phase transition, rather it is just only a gradual change for the finite particle system and the analyses with small number of particles is in support with this argument \cite{JOSE}. However, the higher order corrections become crucial for the systems with isotropic traps and small number of particles. As the anisotropy of the trap and the number of particles increases, the higher order corrections become negligible \cite{JOSE}. In our present consideration of the experimental study, the anisotropy and the number of particles are considerably high, and thus we bother only about first order correction.\\

The transition temperature with the first order correction of the finite size effect can be written as \cite{HOL,KERTLE},
 \begin{equation}\label{eq:1.17}
T^{1}_{c}=T^{0}_{c}-\frac{\zeta(2)}{2\zeta(3)}\frac{\hbar\overline{\omega}}{k_{B}}
\end{equation}
The shift in the transition temperature due to the interaction is given as \cite{ARNO,KASHU},
 \begin{equation}\label{eq:1.18}
\Delta T_{c}= T^{int}_{c}-T_{C}^{nonint}= c a_{sc}n^{1/3}T_{C}^{nonint}
\end{equation}
Where $T^{int}_{c}$ is the transition temperature for the interacting case, $T_{C}^{nonint}$ is the transition temperature for the ideal case, $c=1.32$ is a constant \cite{ARNO}, $a_{sc}$ is the scattering length and $n$ is the density. Thus the transition temperature of the non-interacting case can be represented in terms of transition temperature of the interacting case as,
\begin{equation}\label{eq:1.19}
T_{C}^{nonint}= \frac{T^{int}_{c}}{(1+c a_{sc}n^{1/3})}
\end{equation}
with the scattering length $a_{sc}\approx4.8668\times10^{-9} m$ \cite{DAVI} and the other parameters as we have mentioned above, we calculate $T^{1}_{c}=1.3608261888748566 \mu K$ and the $T_{C}^{nonint}=1.9339862262562904 \mu K$. For the ideal case $T^{1}_{c}$ should be equal to $T_{C}^{nonint}$. At this point we are not making any comment on this difference, but in the course of this paper we will do it.\\

Another important point that one need to note is that the calculated value of $T^{1}_{c}$ is less than the experimentally obtained value $2\mu K$ for the transition temperature. This shows the dominance of effect of the correlation between the particle at the critical point \cite{ARNO,KASHU,BIJ,BAY1,HOLZ1,HOLZ2,KLEI,REP,BAY2,AND,HOLZ3} over the opposing mean field effect which induces a negative shift to the transition temperature \cite{GIO}. Atomic gas of Sodium \cite{EXPARA} is a strongly interacting bose gas, thus the present analyses is in support with the recent high precision measurement study to understand the effects of interactions on the critical temperature of the harmonically trapped atomic bose gas \cite{SMI}, where as the similar experiments given in references \cite{SMI1,SMI2,SMI3} supports the mean-field theory. Ref.\cite{SMI} shows that for the weak interaction the shift in the transition temperature is negative, as in the mean-field theory and for the strong interactions the shift is positive which is the characteristics of critical correlation. However, experimental case \cite{EXPARA} that we have considered for the present study was remained unaddressed by Ref.\cite{SMI}.\\

\begin{figure}
\begin{center}
\includegraphics[height=2.5in,width=3.5in]{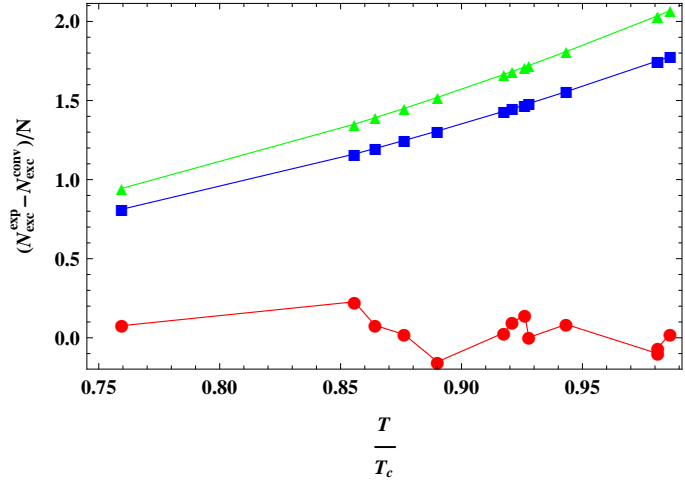}
\caption{Difference in $N_{exc}/N$ for experimental and conventional theoretical value versus $T/T_{c}$. The green filled triangles and the blue filled rectangles represent the first term of the R.H.S and the red filled circles are the L.H.S of the equation (\ref{eq:1.20}). Transition temperature for the conventional theory $T^{conv}_{c}=1.3608261888748566 \mu K$, experimental transition temperature for the interacting case $T^{exp}_{C}=2 \mu K$ (green filled triangles) and experimental transition temperature for the non-interacting case $T^{exp}_{C}=1.9339862262562904 \mu K$ (blue filled rectangles).}\label{fig:1}
\end{center}
\end{figure}

From equation (\ref{eq:1.14}), we can see that the number of particles in the excited states $N_{exc}$ depends on $\beta$. Now in the present study, we claim that the $\beta$ is also a factor that induces a shift in the transition temperature. We consider the generalized expression given in equation (\ref{eq:1.14}) represents the experimental value. Then the difference between the excited fraction of experimental and the conventional theory  (equation (\ref{eq:1.13}) represents the $N_{exc}$ in the conventional theory) can be written as,
\begin{eqnarray}\label{eq:1.20}
\frac{N_{exc}^{exp}-N_{exc}^{conv}}{N}\approx\frac{\zeta(3)}{N}\frac{k_{B}^{3}}{\hbar^{3}\omega_{x}\omega_{y}\omega_{z}}(T_{exp}^{3}-T_{conv}^{3})-\frac{\beta\zeta(3)}{N}\frac{\hbar^{3}\omega_{x}\omega_{y}\omega_{z}}{7k_{B}^{4}T_{exp}^{4}}
\end{eqnarray}
Where $N_{exc}^{exp}$  is the experimentally determined number of particles in the excited states and $N_{exc}^{conv}$ is the number of particles in the excited states calculated from the conventional theoretical relation given in equation (\ref{eq:1.13}). Here, $T_{conv}$ is the $T_{c}^{1}$ given in equation (\ref{eq:1.17}).
We have plotted L.H.S and the first part of the R.H.S for the experimental parameters discussed above and the results are given in Fig.\ref{fig:1}.
The green and the red curves given in Fig.\ref{fig:1} should merge as per the conventional theory represented by equation (\ref{eq:1.13}). Fig.\ref{fig:1} clearly indicates the problem with the difficulty in experimental determination of transition temperature. Not only a particular experiment that we are considering \cite{EXPARA}, but also in all the BEC experiments reported so far have the same problem. This comes because, in experiments actually we plot the experimentally obtained condensate fraction and then fit the curve for the traditional theoretical expression as given in equation (\ref{eq:1.13}). While doing this we are fixing the phase transition to a particular point at which the condensate fraction increases notably and this can be clearly seen in any experimental condensate fraction plot reported so far \cite{JOSE}. This method of fitting will work for the thermodynamic limit where we can talk about a well defined phase transition. However, in the real laboratory experiments the number of particle are finite, thus the phase transition temperature will not be a particular single value, rather it will spread out to a very small interval and this can be noticeable when the number of particles are very less \cite{JOSE}. Thus the experimental studies of BEC have loop holes in it's schemes itself and so one cannot talk about the experimentally determined transition temperature with hundred percentage surety.\\

\begin{figure}
\begin{center}
\includegraphics[height=2.5in,width=3.5in]{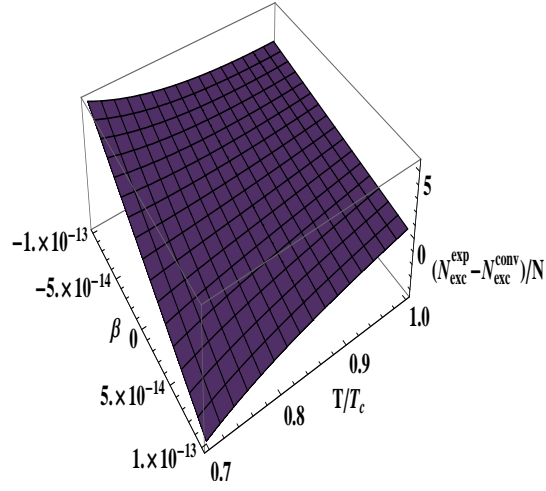}
\caption{For the ideal bose gas, difference in $N_{exc}/N$ for experimental and conventional theoretical value corresponding to $T/T_{c}$ and different $\beta$ value. The $T/T_{c}$ starts from $0.7$, so that we can compare it with Fig.\ref{fig:1}. }\label{fig:2}
\end{center}
\end{figure}

\begin{figure}
\begin{center}
\includegraphics[height=2.5in,width=3.5in]{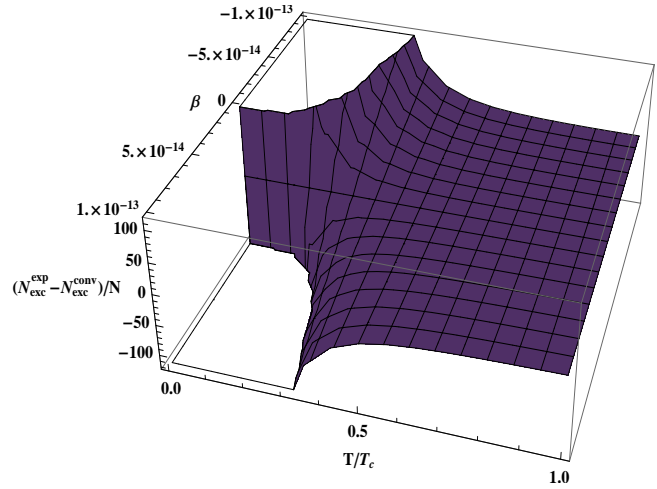}
\caption{For the ideal bose gas, difference in $N_{exc}/N$ for experimental and conventional theoretical value corresponding to $T/T_{c}$ and different $\beta$ value.}\label{fig:3}
\end{center}
\end{figure}

Coming back to the present problem, if we consider the second term in the R.H.S of equation (\ref{eq:1.20}) without $\beta$, it's numerical value is approximately $10^{13}$. That means the second term in the R.H.S is comparable with the first term of the R.H.S and the L.H.S only when $|\beta|\leq10^{-13}$ otherwise, it could have been reflected as a huge variation in the experimental results. Even though we cannot calculate the exact value for $\beta$, we can surely say that $|\beta|\leq10^{-13}$ for this particular experiment. We plot 3D graphs for L.H.S corresponding to the changes in both the $T/T_{c}$ and $\beta$ and is given in Fig.\ref{fig:2} and Fig.\ref{fig:3}. Fig.\ref{fig:2} is  actually a portion of Fig.\ref{fig:3} for the $T/T_{c}$ starts from $0.7$ which is comparable with the two dimensional plot given in Fig.\ref{fig:1}.
\\

In Fig.\ref{fig:2} we can see that as $|\beta|$ becomes very small, we cannot distinguish the presence of second term in the R.H.S of equation (\ref{eq:1.20}). As we have mentioned above, for finite number of particles we do not have any specific transition temperature, rather it will be a small temperature range so that there will always be an error associated with the transition temperature and this small error could possibly suppress the effect of $|\beta|$ if it is very small. However, Fig.\ref{fig:3} shows that as $T\rightarrow0$, irrespective of the value of $|\beta|$, the second term in the equation (\ref{eq:1.20}) dominates and this may be detectable in the experiment also.\\

At the transition temperature, equation (\ref{eq:1.14}) can be written as,
\begin{equation}\label{eq:1.21}
N\approx\zeta(3)(\frac{(k_{B}T_{c})^{3}}{\hbar^{3}\omega_{x}\omega_{y}\omega_{z}}-\beta\frac{\hbar^{3}\omega_{x}\omega_{y}\omega_{z}}{7k_{B}^{4}T_{c}^{4}})
\end{equation}
Where $N$ is the total number of particles. Equation (\ref{eq:1.20}) shows that the $\beta$ will also induce a shift to the transition temperature, but the shift purely depends on the value of $\beta$. From equation (\ref{eq:1.21}), we can write $\beta$ as,
\begin{equation}\label{eq:1.22}
\beta=7\hbar\overline{\omega}((\frac{k_{B}T_{c}}{\hbar\overline{\omega}})^{7}-\frac{N}{\zeta(3)}(\frac{k_{B}T_{c}}{\hbar\overline{\omega}})^{4})
\end{equation}
we can rewrite equation (\ref{eq:1.22}) as,
\begin{equation}\label{eq:1.23}
\beta=7\hbar\overline{\omega}(\frac{N}{\zeta(3)})^{7/3}(x^{7}-x^{4})
\end{equation}
where $x=T_{c}^{nonint}/T_{c}^{0}$. $T_{c}^{nonint}$ is the experimentally observed transition temperature for the ideal bose gas at thermodynamical limit. In the present study, if we assume that there is only interaction effect on the transition temperature, we have $1.3608261888748566\mu K\leq T_{c}^{nonint}\leq 1.9339862262562904\mu K$ and thus $0\leq \beta \leq 1.5672984235789861\times10^{-13}$. This region for $\beta$ is included in the Fig.\ref{fig:2} and Fig.\ref{fig:3}.
\section*{Conclusions}
We have presented the generalized canonical partition function in the frame of standard statistical mechanics and employed it for the study of Bose-Einstein condensation. The generalized canonical partition function has an extra term with a temperature independent constant $\beta$ and we have analysed the BEC with the two different cases of $\beta$, ie. $\beta>0$ and $\beta<0$. For $\beta>0$ the system will become fully condensed one at some temperature $T_{0}$ above absolute zero and will immediately collapse. For $\beta<0$ the system with a partially condensed bose gas will collapse at some positive temperature above absolute zero. This implies that, irrespective of the sign of $\beta$ we cannot have a trapped bose gas, whether it is a partially or fully does not matter at absolute zero. This contradict with the basic theoretical concept of BEC. In fact the huge fluctuation in the ground state occupation in the grand canonical ensemble may be an indication of this.\\

We have analysed the present study with an experimental results. With the experimental details that are available to date, we cannot precisely determine the value of $\beta$, but can give a possible range for the value of $\beta$. We have pointed out that the error produced in the experimental determination of transition temperature due to the finiteness in the number of particle and the biased approach of experimental procedures towards the traditional BEC theory could possibly suppress the effect produced by the second term in the generalized canonical partition, so that it is not detected so far. No matter how small $\beta$ is, at absolute zero we can definitely find its effect.\\

In the present study we have particularly concentrated on ideal bose gas confined in a 3D harmonic potential and thus future studies with the other potential traps can provide more insight towards the problem. However, considering this as an introduction towards a new topic, we demand 'unbiased' experimental studies with ideal bose gas at the thermodynamic limit at absolute zero.
\subsection*{Acknowledgement}
One of the author Sarath would like to thank Dr. Bobby. K. Antony, Dr. Bhalamurugan Sivaraman, Dr. K. P. Subramanian  and the entire Hadron Physics group of the department of Physics, Saradar Patel University for their help and support.

\end{document}